\documentclass[12pt]{article}
\usepackage{amsmath,amssymb}
\begin{document}
\title{{\bf A Unified Convergence Theorem}
\thanks{This Research Is Supported by Natural Science Fund of Zhejiang Province of China (M103057).}}
\author{Junde Wu\\\small \it Department of Mathematics, Zhejiang University, Hangzhou 310027, China\\\small \it
 E-mail: wjd@math.zju.edu.cn
\\Jianwen Luo\\\small \it School of Management, Shanghai Jiao Tong
University, Shanghai 200052, China\\Shijie Lu\\\small \it City
College, Zhejiang University, Hangzhou 310015, China.}
\date {}
\maketitle \thispagestyle{myheadings} \markright{A Convergence
Theorem}

{\bf Abstract.} We prove a unified convergence theorem, which
presents in four equivalent forms of the famous Antosik-Mikusinski
Theorems. In particular, we show that Swartz' three uniform
convergence principles are all equivalent to the
Antosik-Mikusinski Theorems.



\vskip 0.1 in

\vskip 0.1 in \centerline  {\bf 1. Antosik-Mikusinski Theorems}

\vskip 0.1 in

In the late 1950's, J. Mikusinski and R. Sikorski established a
sequential theory of distributions [1, 2]. In order to prove the
equivalence of the sequential approach to the theory of
distributions of Mikusinski and Sikorski and the functional
analysis approach by L. Schwartz [3], Mikusinski noticed that it
would be necessary to develop a new non-topological method
replacing the Baire category methods employed so often in
functional analysis. Mikusinski [4] presented a {\it diagonal
theorem} concerning the diagonal of an infinite matrix with valued
in a Banach space, which proved to be just the tool needed. The
Mikusinski's diagonal theorem indicated that an important infinite
matrix method, which is a non-topological method, came into being.
See [5, $P_{1-7}$] for the evolution of the Mikusinski diagonal
theorem. In 1980's, P. Antosik proved a new diagonal theorem which
was of a different character from the previous diagonal theorems.
Since the appearance of Antosik theorem the infinite matrix method
has taken great progress. In the famous monograph [6] Antosik and
Swartz systematically used this theorem to treat a wide variety of
topics in measure theory and functional analysis. From now on, we
call this theorem the Antosik-Mikusinski Theorem, which states as
follows:

Let $(G, \tau)$ be an Abelian topological group and $x_{ij}\in G$ for
$i, j\in\bf N$. Suppose

(v)  $\lim_ix_{ij}=x_j$ exists for each $j\in\bf N$ and

(vi) for each strictly increasing sequence of positive integers
$\{m_j\}$ there is a subsequence $\{n_j\}$ such that
$\{\sum_jx_{in_j}\}_{i=1}^{\infty}$ is a $\tau$-Cauchy sequence.

Then $\lim_ix_{ij}=x_j$ uniformly for $j\in \bf N$. In particular,
$\lim_ix_{ii}=0$.

However, the Antosik-Mikusinski Theorem is not suitable when we
study the weakly sequentially completeness of the $\beta$-dual
spaces of sequence spaces with the signed weak gliding hump
property. In order to extend the applicability of the
Antosik-Mikusinski Theorem one has to try to establish a new type
of the Antosik-Mikusinski Theorem. Stuart [7, 8] succeeded in
obtaining two such results, which we call the {\it signed}
Antosik-Mikusinski Theorem and the {\it isometry}
Antosik-Mikusinski Theorem, respectively. The above theorems can
be stated as follows:

{\bf The Signed Antosik-Mikusinski Theorem}. Let $(G, \tau)$ be an
Abelian topological group and $x_{ij}\in G$ for $i, j\in\bf N$.
Suppose

 (v) $\lim_ix_{ij}=x_j$ exists for each $j\in\bf N$ and

 (vi) for each strictly increasing sequence of positive integers $\{m_j\}$ there is a subsequence $\{n_j\}$
 and a signed sequence $\{\theta_j\}\subseteq \{1, -1\}$ such that
$\{\sum_{j=1}^{\infty}\theta_j(x_{in_j})\}_{i=1}^{\infty}$ is
Cauchy.

Then $\lim_ix_{ij}=x_j$ uniformly for $j\in \bf N$. In particular,
$\lim_ix_{ii}=0$.

{\bf The Isometry Antosik-Mikusinski Theorem}. Let $(G, P)$ be an
Abelian quasi-normed group and $x_{ij}\in G$ for $i, j\in\bf N$.
Suppose

 (v)  $\lim_ix_{ij}=x_j$ exists for each $j\in\bf N$,

 (vi) for each strictly increasing sequence of positive integers $\{m_j\}$
 there is a subsequence $\{n_j\}$
 and a sequence of additive isometries $\{s_j\}: (G, P)\rightarrow (G,
P)$ such that
$\{\sum_{j=1}^{\infty}s_j(x_{in_j})\}_{i=1}^{\infty}$ is Cauchy.

Then $\lim_ix_{ij}=x_j$ uniformly for $j\in \bf N$. In particular,
$\lim_ix_{ii}=0$.

Being related to these theorems Swartz [5] described a series of
applications of the above three Antosik-Mikusinski Theorems which
did not appear in [6]. Recently, Wu junde and Lu Shijie [9, 10]
also obtained some interesting conclusions from Stuart's results.

\vskip 0.1 in

\centerline  {\bf 2. Uniform Convergence Principles}

\vskip 0.1 in

Let $G$ be an Abelian group. We recall that a functional $P$ on
$G$ is a {\it quasi-norm} if for each $x,y\in G, P(0)=0,
P(-x)=P(x)$ and $P(x+y)\leq P(x)+P(y)$. If $(G, \tau)$ is an
Abelian topological group, then the topology $\tau$ can be
generated by a family of quasi-norms. See Weber [11] for this.

Let $(G, P)$ be an Abelian quasi-normed group. Then $s: (G,
P)\rightarrow (G, P)$ is said to be an {\it additive isometry}, if for
any $x,y\in G$ we have $s(x+y)=s(x)+s(y)$ and $P(s(x))=P(x)$. It
is clear that the identity mapping and $s(x)=-x$ are both additive
isometry.

Let $\Omega$ be a non-empty set and let $(G, P)$ be an Abelian
quasi-normed group. Also, let $\cal F$ be a family of functions which
maps $\Omega$ into $(G, P)$. A sequence $\{f_k\}$ in $\cal F$
is said to be {\it pointwise isometry $\cal K$-convergent} (with respect
to $\cal F$) if for each subsequence $\{f_{m_k}\}$ of $\{f_k\}$
there is a further subsequence $\{f_{n_k}\}$ of $\{f_{m_k}\}$, a
sequence of additive isometries $\{s_k\}$, and a function $f\in
\cal F$ such that $\sum_{k=1}^{\infty}s_k(f_{n_k}(t))=f(t)$ for
each $t\in \Omega$.

If $(G, \tau)$ is an Abelian topological group, we say that
$\{f_k\}\subseteq \cal F$ is {\it pointwise signed $\cal
K$-convergence} (with respect to $\cal F$) if for each subsequence
$\{f_{m_k}\}$ of $\{f_k\}$ there is a further subsequence
$\{f_{n_k}\}$ of $\{f_{m_k}\}$, a signed sequence
$\{\theta_k\}\subseteq \{1, -1\}$, and a function $f\in \cal F$
such that $\sum_{k=1}^{\infty}\theta_kf_{n_k}(t)=f(t)$ for each
$t\in \Omega$.

Similar, we can define the pointwise $\cal K$-convergence
sequences.

A subset $B\subseteq \Omega$ is said to be $\cal F$ {\it sequentially
conditionally compact}, if for each sequence $\{t_j\}\subseteq B$
has a subsequence $\{t_{n_j}\}$ of $\{t_j\}$ such that
$\lim_jf(t_{n_j})$ exists for each $f\in \cal F$.

Let $(\Omega, \tau_1)$ be a Hausdorff topological space. Then we denote by
$C_G(\Omega)$ the space of all continuous functions
$f:(\Omega, \tau_1) \rightarrow (G, P)$.

Also, Swartz [5] proved the following three uniform convergence
principles:

\medskip
{\bf Uniform Convergence Principle I}. Let $(\Omega, \tau_1)$ be a
sequentially compact Hausdorff topological space. Also, let $(G,
\tau)$ be an Abelian topological group and ${\cal F}\subseteq
C_G(\Omega)$. If $\{f_j\}$ in $\cal F$ is pointwise $\cal
K$-convergent (with respect to $\cal F$), then $\{f_j\}$
convergent to 0 uniformly on $\Omega$.

\medskip
{\bf Uniform Convergence Principle II}.  Let $(\Omega, \tau_1)$ be
a compact Hausdorff topological space. Also, let $(G, P)$ be an
Abelian quasi-normed group and ${\cal F}=C_G(\Omega)$. If
$\{f_j\}$ in $\cal F$ is pointwise $\cal K$-convergent (with
respect to $\cal F)$, then $\{f_j\}$ convergent to 0 uniformly on
$\Omega$.

\medskip
{\bf Uniform Convergence Principle III}. Let $\Omega$ be a
non-empty set, $(G, \tau)$ be an Abelian topological group. Also,
let $\cal F$ be a family of functions which maps $\Omega$ into
$(G, \tau)$. If $\{f_j\}$ in $\cal F$ is pointwise $\cal
K$-convergent (with respect to $\cal F$), and $\Omega$ is $\cal F$
sequentially conditionally compact, then $\{f_j\}$ convergent to 0
uniformly on $\Omega$.

\medskip
In next section we prove that the three uniform convergence
principles of Swartz are all equivalent to the Antosik-Mikusinski
Theorem. In particular, we can extend the compact spaces in
Uniform Convergent Principle II to countable compact spaces.

\medskip
Note that, in 1992, Li Ronglu [12] had already established a
uniform convergence principle which is stronger than the UCP (I)
of Swartz ([5], 1996). Using this result, Li Ronglu studied the
summability of a class of operator matrices in [13]. Also, Qu
Wenbo and Wu Junde [14] pointed out that the uniform convergence
principle of Li Ronglu [12] is equivalent to the
Antosik-Mikusinski Theorem.

\vskip 0.1 in \centerline  {\bf 3. The Main Theorem and Its Proof}
\vskip 0.1 in

At first, we need to generalize the Antosik Lemma of [15].

{\bf Lemma 1}. Let $G$ be an Abelian quasi-normed group and $x_{ij
}\in G$, $i, j\in\bf N$. Suppose that for each strictly increasing
sequence $\{m_i\}$ in $\bf N$ has a subsequence $\{n_i\}$ and a
sequence of additive isometries $s_j: (G, P)\rightarrow (G, P)$
such that

   (i) $\lim_ix_{n_in_j}=0$ for each $j\in \bf N$,

   (ii) $\lim_i\sum_{j=1}^{\infty}s_j(x_{n_in_j})=0$.

 Then $\lim_i x_{i i}=0$ .

{\bf Proof}. If not, there is an increasing sequence of positive
integers $\{m_i\}$ and $\varepsilon_0>0$ such that for each $i\in
\bf N$, $$P(x_{m_im_i})\geq\varepsilon_0. \eqno(1)$$ Let $\{n_i\}$
be a subsequence of $\{m_i\}$ such that (i) and (ii) hold. It is
clear that for $i\in\bf N$ and $j\rightarrow \infty,
P(x_{n_in_j})\rightarrow 0$; for $j\in\bf N$ and $i\rightarrow
\infty, P(x_{n_in_j})\rightarrow 0$. Put $k_1=1$ and find an index
$k_2$ such that $P(x_{n_{k_i}n_{k_j}})<2^{-i-j}$ for $i, j=1, 2$
and $i\not=j$. By induction we can find an increasing sequence
$\{k_i\}$ such that $P(x_{n_{k_i}n_{k_j}})<2^{-i-j}$ for $i,
j\in\bf N$ and $i\not=j$. Putting $p_i=n_{k_i}$ we have
$$P(x_{p_ip_j})<2^{-i-j}$$ for $i,j\in \bf N$ and $i\not=j$.
Let $\{q_i\}$  be a subsequence
of $\{p_i\}$ and let $\{s_j\}$ be a sequence of additive isometries of
$(G, P)\rightarrow (G, P)$ such that
$$\lim_i\sum_{j=1}^{\infty}s_j(x_{q_iq_j})=0.$$
Note that $$\sum_{j\not= i}P(s_j(x_{q_iq_j}))=\sum_{j\not=
i}P(x_{q_iq_j})<2^{-i},$$
$$P(\sum_{j=1}^{\infty}s_j(x_{q_iq_j}))\rightarrow 0,$$ and
$$P(x_{q_iq_i})=P(s_i(x_{q_iq_i}))\leq \sum_{j\not=
i}P(s_j(x_{q_iq_j}))+P(\sum_{j=1}^{\infty}s_j(x_{q_iq_j}))$$ for
$i\in \bf N$. Hence we have $P(x_{q_iq_i})\rightarrow 0$. This
contradicts (1), which proves the lemma.

The following lemma is also needed.

{\bf Lemma 2 [16].} Let $(X,\tau_1)$ be a countably compact
Hausdorff topological space and let $(Y,d)$ be a metric space.
Then each continuous bijection $T: (X,\tau_1)\rightarrow (Y,d) $
is a homeomorphism.

Our main result is:

{\bf Theorem 1.} Let $(G, P)$ be an Abelian quasi-normed group.
The following six statements are equivalent and they all hold true:

\medskip
(I). If for $i, j\in{\bf N}, z_{ij}\in G$, and

(iii) $\lim_iz_{ij}=0$ for each $j\in\bf N$,

(iv) For each strictly increasing sequence of positive integers
$\{m_j\}$ there is a subsequence $\{n_j\}$ of $\{m_j\}$ and a
sequence of additive isometries $\{s_j\}: (G, P)\rightarrow (G,
P)$ such that $\lim_i\sum_{j=1}^{\infty}s_j(z_{in_j})=0$.

Then $\lim_iz_{ii}=0$.

\medskip
 (II). (The Isometry Antosik-Mikusinski Theorem). If for $i, j\in{\bf N}, x_{ij}\in G$, and

 (v)  $\lim_ix_{ij}=x_j$ exists for each $j\in\bf N$,

 (vi) for each strictly increasing sequence of positive integers $\{m_j\}$ there is a subsequence $\{n_j\}$
 and a sequence of additive isometries $\{s_j\}: (G, P)\rightarrow (G,
P)$ such that
$\{\sum_{j=1}^{\infty}s_j(x_{in_j})\}_{i=1}^{\infty}$ is Cauchy.

Then $\lim_ix_{ij}=x_j$ uniformly for $j\in \bf N$. In particular,
$\lim_ix_{ii}=0$.

\medskip
(III). Let $\Omega$ be a non-empty set and $\cal F$ be a family of
functions which mapping $\Omega$ into $(G, P)$. If the sequence
$\{f_i\}$ in $\cal F$ is pointwise isometry $\cal K$-convergent
(with respect to $\cal F$) and $\Omega$ is $\cal F$ sequentially
conditionally compact, then $\{f_i\}$ converges to 0 uniformly on
 $\Omega$.

\medskip
 (IV). Let $(\Omega, \tau_1)$ be a sequentially compact Hausdorff space and $\cal F$ be a
family of continuous functions which mapping $(\Omega, \tau_1)$
into $(G, P)$. If $\{f_i\}$ is pointwise isometry $\cal
K$-convergent (with respect to $\cal F$), then $\{f_i\}$ converges
to 0 uniformly on $\Omega$.

\medskip
(V). Let $(\Omega, \tau_1)$ be a countable compact Hausdorff space
and $\cal F$ be a family of continuous functions which mapping
$(\Omega, \tau_1)$ into $(G, P)$. If $\{f_i\}$ is pointwise
isometry $\cal K$-convergent (with respect to $\cal F$), then
$\{f_i\}$ converges to 0 uniformly on $\Omega$.

\medskip
(VI). Let $(\Omega, \tau_1)$ be a compact Hausdorff space and
$\cal F$ be a family of continuous functions which mapping
$(\Omega, \tau_1)$ into $(G, P)$. If $\{f_i\}$ is pointwise
isometry $\cal K$-convergent (with respect to $\cal F$), then
$\{f_i\}$ converges to 0 uniformly on $\Omega$.
\medskip

{\bf Proof.} (I)$\Rightarrow$ (II). If the conclusion fails, there
are a $\varepsilon_0>0$ and two strictly increasing sequences of
positive integers $\{p_k\}$ and $\{q_k\}$ such that
$$P(x_{p_kq_k}-x_{q_k})\geq \varepsilon_0\eqno(2)$$ for all $k\in \bf N$. Note
that $$x_{p_iq_j}-x_{q_j}\rightarrow 0$$ for each $j\in \bf N$ as
$i\rightarrow \infty$, therefore, there exists a subsequence
$\{m_i\}$ of $\{p_i\}$ such that
$$P(x_{m_iq_i}-x_{q_i})<\frac{\varepsilon_0}{2}\eqno(3)$$ for $i\in
\bf N$. On the other hand, we have
$$x_{p_iq_i}-x_{q_i}=(x_{p_iq_i}-x_{m_iq_i})+(x_{m_iq_i}-x_{q_i}). \eqno(4)$$

Consider the infinite matrix $(x_{p_iq_j}-x_{m_iq_j})_{ij}$ and
note that the matrix satisfies condition (I). Consequence
$$x_{p_iq_i}-x_{m_iq_i} \rightarrow 0 $$ as $i\rightarrow \infty$,
and
$$P(x_{p_iq_i}-x_{m_iq_i})<\frac{\varepsilon_0}{2}$$ for sufficiently large
$i$. Hence, by (4) and (3) $$
P(x_{p_iq_i}-x_{q_i})<\varepsilon_0$$ for sufficiently large $i$.
Which contradicts (2) and (I) $\Rightarrow$ (II) holds.

\medskip
(II)$\Rightarrow$ (III). We are to show that
$f_j(\omega)\rightarrow 0$ uniformly for $\omega$ in $\Omega$ or,
equivalently, for each sequence $\{\omega_i\}$ in $\Omega$,
$$f_i(\omega_i)\rightarrow 0 \eqno (5)$$ as $i\rightarrow \infty$.
Let $\{\omega_i\}$ be a sequence of $\Omega$. Since $\Omega$ is
$\cal F$ sequentially conditionally compact, there exists a
subsequence $\{\omega_{n_i}\}$ of $\{\omega_i\}$ such that for
each $f\in \cal F$, $\lim_if(\omega_{n_i})$ exists. Consider the
infinite matrix $(f_{n_j}(\omega_{n_i}))_{ij}$. Note that
$\{f_i\}$ is pointwise isometry $\cal K$-convergent, so the matrix
satisfies condition (II). Thus $$f_{n_i}(\omega_{n_i})\rightarrow
0.$$ Since the same argument can be applied to any subsequence of
$\{f_i(\omega_i)\}$, it follows that  $\lim_if_i(\omega_i)=0$.
(II) $\Rightarrow$ (III) is true.

\medskip
(III)$\Rightarrow$ (IV) and (V) $\Rightarrow $(VI) are clear.

\medskip
(IV)$\Rightarrow$(V). Define an equivalent relation $\sim$ on
$\Omega$ by $t\sim s\Leftrightarrow f_j(t)=f_j(s)$ for all $j\in
\bf N$. Let $\hat{t}$ be the equivalence class determined by $t\in
\Omega$ and let $\hat{\Omega}=\{\hat{t}: t\in\Omega\}$ be the set
of all equivalence classes. Define a metric $d$ on $\hat{\Omega}$
by $$d(\hat{t}, \hat{s})=\sum_{j=1}^\infty{1\over
2^j}{{P(f_j(s)-f_j(t))}\over{1+ P(f_j(s)-f_j(t))}}.$$ Since each
$f_i$ is a continuous function which mappings $\Omega$ into $(G,
P)$, so map $t\rightarrow \hat{t}$ from $\Omega$ onto
$\hat{\Omega}$ is a continuous bijection with respect to $d$. It
follows from Lemma 2 that $t\rightarrow \hat{t}$ is a
homeomorphism. So the quotient topology of $\hat{\Omega}$ is
equivalent to the metric topology $d$. Thus, $(\hat{\Omega}, d)$
is a countable compact metric space, so $(\hat{\Omega}, d)$ is a
sequentially compact metric space. Define  $\hat{f_i}:
\hat{\Omega}\rightarrow G$ by $\hat{f_i}(\hat{t})=f_i(t)$. Note
that each $\hat{f_i}$ is well-defined and continuous. Let $\cal F$
be the all continuous functions which mapping $\hat{\Omega}$ into
$G$. We claim that $\{\hat{f_i}\}$ is pointwise isometry $\cal
K$-convergent with respect to $\cal F$. Given any subsequence of
$\{\hat{f_i}\}$, there is a further subsequence
$\{\hat{f_{n_i}}\}$, a sequence of additive isometries $\{s_j\}:
(G, P)\rightarrow (G, P)$ and a continuous function $f$ which
mappings $\Omega$ into $G$ such that
$\sum_{i=1}^{\infty}s_i(f_{n_i}(t))=f(t)$ for all $t\in \Omega$.
Define $\hat{f}: \hat{\Omega}\rightarrow G$ by
$\hat{f}(\hat{t})=f(t)$. Note that $\hat{f}$ is well-defined since
if $t\sim s$, then $f_{n_i}(t)=f_{n_i}(s)$ for each $i\in \bf N$.
Furthermore, $\hat{f}$ is a continuous function which mappings
$(\hat{\Omega}, d)$ into $(G, P)$ and have
$\sum_{i=1}^{\infty}s_i(\hat{f_{n_i}}(\hat{t}))=\hat{f}(\hat{t})$
for all $t\in \Omega$. So $\hat{\{f_i}\}$ is pointwise isometry
$\cal K$-convergent. By (IV) that
$\lim_i\hat{f_i}(\hat{t})=\lim_if_i(t)=0$ uniformly for $t\in
\Omega$. Thus, (IV)$\Rightarrow$ (V) holds.

\medskip
(VI)$\Rightarrow$(I). Indeed, if the hypothesis of (I) is
satisfied. Let $\Omega = \{\frac{1}{i} , 0\}_{i=1}^{\infty} $, for
$x, y\in \Omega$, put $d_1(x, y)=|x-y|$. Then $(\Omega, d_1)$ is a
compact Hausdorff topological space. Let $f_j: \Omega
\longrightarrow G$ satisfies that if $\omega=\frac {1}{i},
f_j(\omega)=z_{ij}$; if $\omega=0, f_j(\omega)=0$. It is easily to
show that each $f_j$ is continuous and for each strictly
increasing sequence $\{m_j\}$ in $\bf N$ has a subsequence
$\{n_j\}$ and a sequence of additive isometries $\{s_j\}: (G,
P)\rightarrow (G, P)$ such that for each $i\in \bf N$, the series
$\sum_js_j(z_{in_j})$ is convergent and
$\lim_i\sum_js_j(z_{in_j})=0$. Thus, for each $\omega\in \Omega$ ,
the series $\sum_js_j(f_{n_j}(\omega))$ is convergent and
$\sum_js_jf_{n_j}: \Omega \longrightarrow G$ is continuous. It
follows from (VI) that $\lim_jf_j(\omega)=0$ uniformly on
$\Omega$. In particular, $\lim_jf_j(\frac{1}{j})=\lim_jz_{jj}=0$.
So (I) is hold.

Thus, we have proved that all six statements (I), (II), (III),
(IV), (V), (IV) are equivalent. Since Lemma 1 $\Rightarrow$(I) is
obviously true and the six statements are equivalent and they all
hold true, thus, we proved the theorem.

Note that Theorem 1 has unified the Isometry Antosik-Mikusinski
Theorem and Swartz's Isometry Type Uniform Convergence Principles.

Since the topology $\tau$ of each Abelian topological group $(G,
\tau)$ can be generated by a family of quasi-norms as in Weber
[11], we can obtain the following signed type analog of Theorem 1
as Corollary:

{\bf Corollary 1}. Let $(G, \tau)$ be an Abelian topological
group. The following states are equivalent and they all hold true:

\medskip
(I). If for $i, j\in{\bf N}, z_{ij}\in G$, and

(iii) $\lim_iz_{ij}=0$ for each $j\in\bf N$,

(iv) for each strictly increasing sequence of positive integers
$\{m_j\}$ there is a subsequence $\{n_j\}$ and a signed sequence
$\{\theta_j\}\subseteq \{1, -1\}$ such that
$\lim_i\sum_{j=1}^{\infty}\theta_jz_{in_j}=0$.

Then $\lim_iz_{ii}=0$.

\medskip
(II). (The Signed Antosik-Mikusinski Theorem). If for $i, j\in{\bf N}, x_{ij}\in G$, and

 (v)  $\lim_ix_{ij}=x_j$ exists for each $j\in\bf N$,

 (vi) for each strictly increasing sequence of positive integers $\{m_j\}$ there is a subsequence $\{n_j\}$
 and a signed sequence $\{\theta_j\}\subseteq \{1, -1\}$ such that
$\{\sum_{j=1}^{\infty}\theta_jx_{in_j}\}_{i=1}^{\infty}$ is
Cauchy.

Then $\lim_ix_{ij}=x_j$ uniformly for $j\in \bf N$. In particular,
$\lim_ix_{ii}=0$.

\medskip
(III). Let $(\Omega, \tau_1)$ be a non-empty set and $\cal F$ be a
family of functions which mapping $\Omega$ into $(G, \tau)$. If
the sequence $\{f_i\}$ in $\cal F$ is pointwise signed $\cal
K$-convergent (with respect to $\cal F$) and $\Omega$ is $\cal F$
sequentially conditionally compact, then $\{f_i\}$ converges to 0
uniformly on
 $\Omega$.

\medskip
 (IV). Let $(\Omega, \tau_1)$ be a sequentially compact Hausdorff space and $\cal F$ be a
family of continuous functions which mapping $\Omega$ into $(G,
\tau)$. If $\{f_i\}$ is pointwise signed $\cal K$-convergent (with
respect to $\cal F$), then $\{f_i\}$ converges to 0 uniformly on
$\Omega$.

\medskip
(V). Let $(\Omega, \tau_1)$ be a countable compact Hausdorff space
and $\cal F$ be a family of continuous functions which mapping
$\Omega$ into $(G, \tau)$. If $\{f_i\}$ is pointwise signed $\cal
K$-convergent (with respect to $\cal F$), then $\{f_i\}$ converges
to 0 uniformly on $\Omega$.

\medskip
(VI). Let $(\Omega, \tau_1)$ be a compact Hausdorff space and
$\cal F$ be a family of continuous functions which mapping
$\Omega$ into $(G, \tau)$. If $\{f_i\}$ is pointwise signed $\cal
K$-convergent (with respect to $\cal F$), then $\{f_i\}$ converges
to 0 uniformly on $\Omega$.
\medskip

Now, we give a direct application of Corollary 1.

Let $(\Omega, \tau_1)$ be a compact Hausdorff space and $(X,
||.||)$ be a normed space. Then an interesting question in
Analysis arises as follows: If a series $\sum_kf_k$ in
$C_X(\Omega)$ convergent to $f\in C_X(\Omega)$ pointwise on
$\Omega$, then under what conditions the series $\sum_kf_k$
convergent to $f$ uniformly on $\Omega$? Thomas in [17] answered
this question:

If $\sum_kf_k$ convergent to $f\in C_X(\Omega)$ pointwise on
$\Omega$, and for each subsequence $\{f_{n_k}\}$ of $\{f_k\}$
there is a $f_0\in C_X(\Omega)$ such that
$\sum_{k=1}^{\infty}f_{n_k}$ convergent to $f_0$ pointwise on
$\Omega$, then $\sum_kf_k$ convergent to $f$ uniformly on
$\Omega$.

Now, we can use Corollary 1 to extend the above Thomas theorem to
more stronger form:

{\bf Corollary 2.} Let $(G, \tau)$ be an Abelian topological
group, $(\Omega, \tau_1)$ be a sequentially compact or countable
compact or, in particular, a compact Hausdorff space,
$\{f_k\}\subseteq C_G(\Omega)$. If for each subsequence
$\{f_{n_k}\}$ of $\{f_k\}$ there is a $f_0\in C_G(\Omega)$ such
that $\sum_{k=1}^{\infty}f_{n_k}(t)=f_0(t)$ for each $t\in
\Omega$, then the series $\sum_{k=1}^{\infty}f_k$ convergent
uniformly on $\Omega$.

Corollary 1 showed that the Thomas theorem of 70's is very close
to the Antosik-Mikusinski Theorem of 80's.

Finally, we point out that the continuity of maps of Theorem 1 and
Corollary 1 can be weaked to the sequentially continuity as was
considered by Li Ronglu in [12, 1992].

\medskip

\vskip 0.2 in

{\bf Acknowledgment.} The first author thanks Prof. Li Ronglu for
his helpful discussion and suggestion on this topics.

\vskip 0.1 in \centerline {\bf REFERENCES }

\noindent [1]. J. Mikusinski, R. Sikorski. The elementary theory
of distributions I. Rozprawy Mat. 12(1957)
\\

\noindent [2]. J. Mikusinski, R. Sikorski. The elementary theory
of distributions II. Rozprawy Mat. 25(1961)
\\

\noindent [3]. L. Schwartz. Theorie des distributions. Hermann,
Paris, 1966
\\

\noindent [4]. J. Mikusinski. A theorem on vector matrices and its
applications in measure theory and functional analysis. Bull.
Acad. Polon. Sci. 18(1970), 193-196
\\

\noindent [5]. C. Swartz. Infinite matrices and the gliding hump,
World Sci. Publ., Singapore, 1996
\\

\noindent [6]. P. Antosik, C. Swartz. Matrix methods in analysis,
Springer Lecture Notes in Math. 1113, Heidelberg, 1985
\\

\noindent [7]. C. Stuart. Weak sequentially completeness in
sequence spaces. Ph. D. Dissertation, New Mexico State University,
1993
\\

\noindent [8]. C. Stuart. Interchanging the limits in a double
series. Southeast Asian Bull. Math. 18(1994), 81-84
\\

\noindent [9]. Junde Wu, Shijie Lu. A summation theorem and its
applications. J. Math. Anal. Appl. 257(2001), 29-38
\\

\noindent [10]. Junde Wu, Shijie Lu. A full invariant theorem and
some applications. J. Math. Anal. Appl. 270(2002), 397-404.
\\

\noindent [11]. H. Weber. A diagonal theorem. Answer to a question
of Antosik. Bull. Polish Acad. Sci. Math., 41(1993), 95-102
\\

\noindent [12]. Ronglu Li, Minhyung Cho. A uniform convergent
principle. J. Harbin Institute of Technology 24(1992), 107-108
\\

\noindent [13]. Ronglu Li, Longsuo Li, Shin Min Kang . Summability
results for operator matrices on topological vector spaces. Sci.
China Ser. A 44(2001), 1300-1311
\\

\noindent [14]. Wenbo Qu, Junde Wu. On Antosik's lemma and the
Antosik-Mikusinski basic matrix theorem. Proc. Amer. Math. Soc.
130(2002), 3283-3285
\\

\noindent [15]. P. Antosik. A lemma on matrices and its
applications. Contemporary Math. 52(1986) 89-95
\\

\noindent [16]. Junde Wu, Ronglu Li, Wenbo Qu. The extension of
Eberlein-Smulian theorem in locally convex spaces, Acta Math.
Sinica, 41(1998), 663-666
\\

\noindent [17]. G.E.F. Thomas. L'Integration par rapport a une
mesure de Radon vectorielle, Ann. Inst. Fourier (Grenoble),
20(1970), 55-191

\end{document}